# Study of simulations of double graded InGaN solar cell structures

Mirsaeid. Sarollahi[a], Manal A. Aldawsari[b], Rohith Allaparthi[a], , Malak A. Refaei[b], Reem Alhelais[b], Md Helal Uddin Maruf [c], Yuriy Mazur[d],
Morgan E. Ware[a,b,c,d]

[a]University of Arkansas, Electrical Engineering Department, 3217 Bell Engineering Center, Fayetteville, AR 72701
[b]University of Arkansas, Microelectronics-Photonics Program, 731 West Dickson Street, Fayetteville, Arkansas 72701, [c]University of Arkansas, Material Science and Engineering 731 West Dickson Street, Fayetteville, Arkansas 72701, [d]Institute for Nanoscience and Engineering, Fayetteville, Arkansas 72701.

**Abstract:** The performances of various configurations of InGaN solar cells are compared using nextnano software. Here we compare a flat base graded wall GaN/InGaN structure, with an $In_xGa_{1-x}N$ well with sharp GaN contact layers, and an $In_xGa_{1-x}N$ structure with $In_xGa_{1-x}N$ contact layers, i.e. a homojunction. The doping in the graded structures are the result of polarization doping at each edge (10 nm from each side) due to the graded structure, while the well structures are intentionally doped at each edge (10 nm from each side) equal to the doping concentration in the graded structure. The solar cells are characterized by their open-circuit voltage, $V_{oc}$, short circuit current, $I_{sc}$, solar efficiency, $\eta$, and energy band diagram. The results indicate that an increase in $I_{sc}$ and $\eta$ results from increasing both the fixed and the maximum indium compositions, while the $V_{oc}$ decreases. The maximum efficiency is obtained for the InGaN well with 60% In.

**Introduction**

The optical properties of InGaN ternary alloys make them interesting materials for photovoltaic devices [1]–[3]. Properties such as direct and tunable bandgaps which cover the whole range of the solar spectrum as well as high thermal conductivity, high optical absorption, and high radiation resistance are center to this interest [4], [5]. As a result, InGaN continues to receive attention in the research community both fundamentally and with a strong push towards photovoltaics [6]–[9],[10]. There have been many theoretical studies of single composition InGaN solar cells. For example, in 2007, modeling optimization of a thick, single junction $In_{0.65}Ga_{0.35}N$ solar cells achieved a conversion efficiency of 20.28%[11]. In another report (in 2008), they obtained higher efficiency (24.95%) with the same indium composition due to adoption of the density of states (DOS) model, providing much more information about recombination/generation in semiconductors than the lifetime model by neglecting defects[12]. These are similar to several other studies of InGaN solar cells, which use different modeling software to report solar efficiency[13].

In the past, graded structures (generally single grades) have been investigated in general for their strain relieving properties[14]. Ternary graded films in III-nitride materials have additionally resulted in polarization doping which has been demonstrated to achieve very high levels of doping, without the use of additional impurities in the lattice[15]. This is explained as follows. InN and GaN have different spontaneous crystal polarizations, *P*. When a compositionally graded structure of InGaN is grown, polarization doping is introduced proportional to the increasing (decreasing) composition resulting in a fixed charge field given by, $-\nabla . P = \rho$. When the compositional grading is from GaN to InGaN for metal polar growth, the background charges are negative ($-\nabla . P = \rho < 0$) which attracts holes to create p-type doping. In the reverse condition, when the grading is from InGaN to GaN for metal polar growth, the background charges are positive ($-\nabla . P = \rho > 0$) which attracts free electrons to create n-type doping[16]. It has been shown that AlGaN polarization charge fields are reversed in comparison with InGaN. In other words, when the alloy is graded from GaN to $Al_xGa_{1-x}N$ for metal polar growth, a positive polarization charge field is created which attracts electrons (n-type), and when the alloy is graded from $Al_xGa_{1-x}N$ to GaN holes are attracted (p-type)[17]. Additionally, for strained films the effects of piezoelectric polarization must be considered similar to the spontaneous polarizations above [18]

Here, we present a simulation study of graded and sharp barrier, GaN/InGaN compared with InGaN single junction solar cells, where the active region is formed by an InGaN layer and the contact layers are either compositionally graded InGaN (polarization doped) or GaN (impurity doped). The structures in this work are shown in Fig. 1 and are all 100 nm thick $In_xGa_{1-x}N$ p-i-n junction structures with three different configurations. The first structure, Fig. 1a, is a flat base-graded barrier structure in which a single layer with constant $x$ is sandwiched between two graded barrier layers. The first graded layer (0-10 nm) starts from 0% to a maximum value, $x_{max}$, then is followed by an 80 nm flat base layer of constant composition, $x_{max}$, and finally (90-100 nm) the composition is graded back to 0%. This is structure A. This results in a flat base graded (FBG) structure with a composition grading up and down at the edges and subsequent polarization doping, which forms a p-i-n junction. In addition, energy band diagram related to structure A (for $x_{max} = 50\%$ at zero volts) is shown at Fig. 1b. The doping concentration resulting from the polarization doping as a function of $x_{max}$ in the graded layers is shown in Fig.1g. The second structure, Fig. 1c, is a square well, with two doped GaN layers at each edge The GaN layers are 10 nm thick each (0-10 nm and 90-100 nm) and the InGaN is 80 nm (10-90 nm) (structure B). Again, the energy band diagram of structure B (for x =50% at zero volts) is displayed in Fig. 1d in which the two edges are doped intentionally, there are two sharp barriers at the edges resulting from a two dimensional sheet charge due to the polarization change at the interface between InGaN and Gan at the beginning (0-10 nm) and at the end (90-100 nm). The third structure is a 100 nm $In_xGa_{1-x}N$ homojunction, with $x = x_{max}$, (structure C). The energy band diagram of structure C (for x =50% at zero volts) is shown in Fig. 1f, which is simply a constant composition InGaN homojunction does not show any clear barrier with the edges doped intentionally. The doping edge layers, 10 nm, in structures B and C are similar in concentration to the comparable $x_{max}$. In all structures (A, B and C) the layers from -10 nm to 0 nm and from 100 nm to 110 nm are considered metal ohmic contacts.

Using nextnano, the band structure is modeled in the envelope function approximation, using single-band effective mass approximation[19]. Generally, the wave function and energy level in a quantum system can be calculated by solving the Schrödinger equation with a suitable Hamiltonian. The electron motion can be explained by a one electron Hamiltonian while multi-body interacts among electrons is considered negligible.

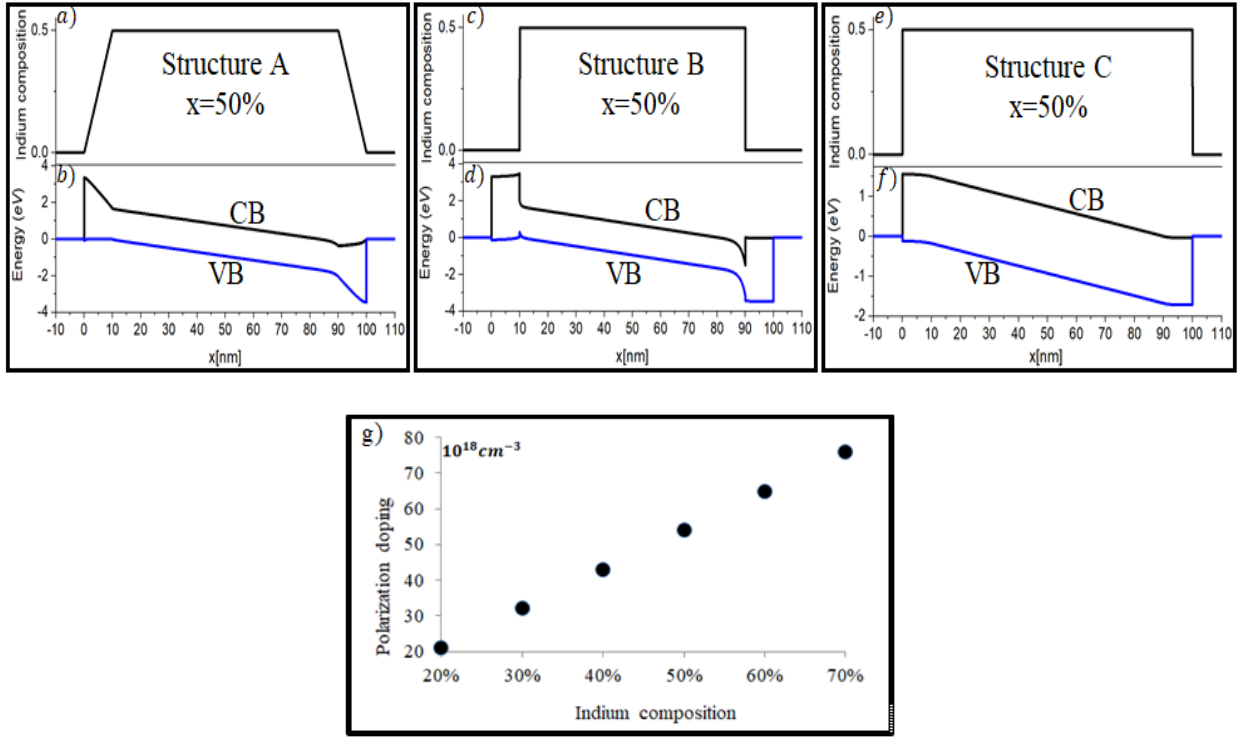

Fig1. a) Flat base Graded triangle well (FBGTW) structure (Structure A) b) energy band diagram for structure A , c) Single square well GaN/InGaN (Structure B) d) energy band diagram for structure B e) Single InGaN homojunction (Structure C) f) energy band diagram for structure C. g) Polarization-doping correlated with indium composition in structure A

**Nextnano Simulation**

In order to obtain solar cell parameters nextnano+ software was used. Nextnano+ only calculates generation rates for fixed composition materials, within a single simulation, so rates for variable composition films must be determined in a step-graded procedure, semi-manually. Total generation versus position can then be imported into the nextnano+ simulation of the full graded structure in order to calculate the light I-V curve and other solar cell parameters.

The FBG contains a $d$ nm graded layer from GaN to a maximum composition of indium, $x_{max}$, followed by a thick layer ($L-2d$) with constant $x_{max}$, and finally another $d$ nm layer graded back to GaN from $x_{max}$. The whole structure of thickness $L$ nm is shown in Fig. 2.

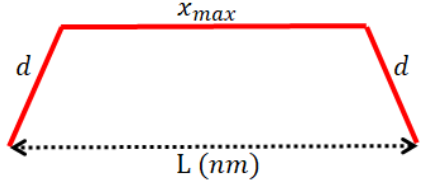

Fig 2. Flat base graded triangle well structure details

To determine the generation rates we discretized the gradient of the material. So, a simple formula was introduced to divide the graded structure to several layers with constant In composition each. A step size determines the thickness of each constant composition layer, such that the In change rate per nm is given by:

$$\frac{x_{max}\%}{d-(step\ size)} \qquad (1)$$

For example, with an $x_{max}$ of 20%, and a step size of 2.5 nm this results in,

$$\frac{20\%}{10-2.5} = 2.67\frac{\%}{nm}$$

This results in a step-graded structure as in Fig. 3.

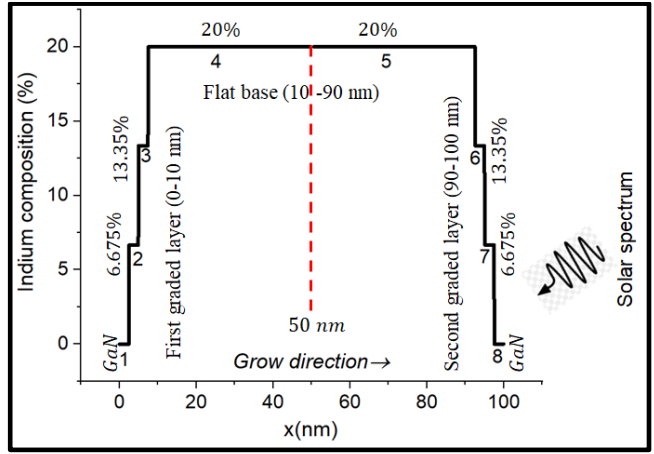

Fig 3. Converting graded structure to step graded by equation 1 to calculate generation rate in nextnano. Receptivity of layers is given by the number below each layer when light passing through the structure.

In order to create the generation rate versus thickness data which is imported into nextnano+ in order to simulate the full solar cell parameters, the generation rate for each of the layers shown in Fig. 3 must be calculated separately. The solar spectrum with intensity, $I_0$, first enters the structure through layer 8 (2.5 nm GaN). The solar spectrum along with the absorption coefficient of layer 8 are imported in nexatnano3, resulting in the generation rate of that layer from 100-97.5 nm. The same procedure is followed for each subsequent layer by considering the light intensity, which is diminished by passing each previous layer. For example the light intensity incident on layer 7 is given by, $I_8 = I_0 \exp(-\alpha_8 d)$, for which $d$ is the thickness of each layer and $\alpha_8$ is absorption coefficient for layer 8. By continuing this procedure for all layers, the generation rate as a function of position within the entire step graded structure is obtained ($G_1(0 - 2.5\ nm), G_2(2.5 - 5\ nm), G_3(5 - 7.5\ nm), G_{4,5}(7.5 - 92.5\ nm), \ldots G_8(97.5 - 100\ nm)$). This is then imported into nextnano and used as the generation rate data for the entire, continuous structure, in order to determine the solar cell parameters.

**Effect of strain on bandgap calculation**

The quality of InGaN devices strongly depends on the substrate used in growth. Due to lattice mismatch between GaN and InGaN strain resulting from the growth of these types of heterostructures can also have a significant impact on the properties and quality of the resulting devices. Heterostructures made of semiconductor materials with different lattice constants are subject to elastic deformations. Such deformations can be studied based on classical elasticity within the harmonic approximation, i.e. for small strains. As lattice deformation varies with the growth direction, a detailed understanding of the strain is important for design, development and study of optoelectronic and electronic devices[20].

Strain causes piezoelectric effects, influences the conduction and valence band edges (including their degeneracies) and the k · p Hamiltonian of Schrodinger equation. Additionally, other parameters correlating with the bandgap are affected like the absorption coefficient. Therefore, strain is a very important parameter for device engineers to modify and control the electronic and optical properties of semiconductor heterostructures [20]. The strain tensor ε(x), i.e. the symmetrical part of the deformation tensor, is defined as:

$$\varepsilon_{ij}(x) = \frac{1}{2}\left(\frac{du_i}{dx_j} + \frac{du_j}{dx_i}\right) = \frac{1}{2}(u_{ij} + u_{ji}) = \varepsilon_{ji} \qquad (2)$$

where $i, j = \{1, 2, 3\}$ and $u(x)$ describes the displacement due to lattice deformations. The strain tensor $\varepsilon$ is symmetric whereas the distortion tensor $u$ is in general not symmetric[20].

In order to model the absorption coefficients of the continuously varying InGaN alloys, the bandgap energies are required. Strain and the resulting piezoelectric polarization and spontaneous polarization are both taken into account in nextnano [21], [22]. So, nextnano was used to predict the bandgap values for $In_xGa_{1-x}N$ strained to a GaN substrate, while Vegard's law, Eq.3 was used in the absence of strain.

$$E_g(In_xGa_{1-x}N) = E_g(InN)x + E_g(GaN)(1-x) - b(x)(1-x), \qquad (3)$$

Large discrepancies are found in the published values of the bowing coefficient, $b$, varying in the range of 1.4eV–2.8eV. Such a broad range of values is generally understood to be the result of poorly defined or measured strain within the $In_xGa_{1-x}N$ layers [23]–[26]. Additionally, there is some controversy about whether there is a compositional dependence in the bowing coefficient [27], [28] or not [13], [29], [30]. Based on the available data, we use a value of $b = 1.43$ eV. According to nextnano database bowing parameter for $In(x)Ga(1-x)N$ is around given value[31]–[33] The two resulting curves showing the bandgap energy as a function of indium composition for both strained (to the GaN lattice constant) and bulk unstrained InGaN are plotted in Fig. 5. Red is unstrained and blue is strained.

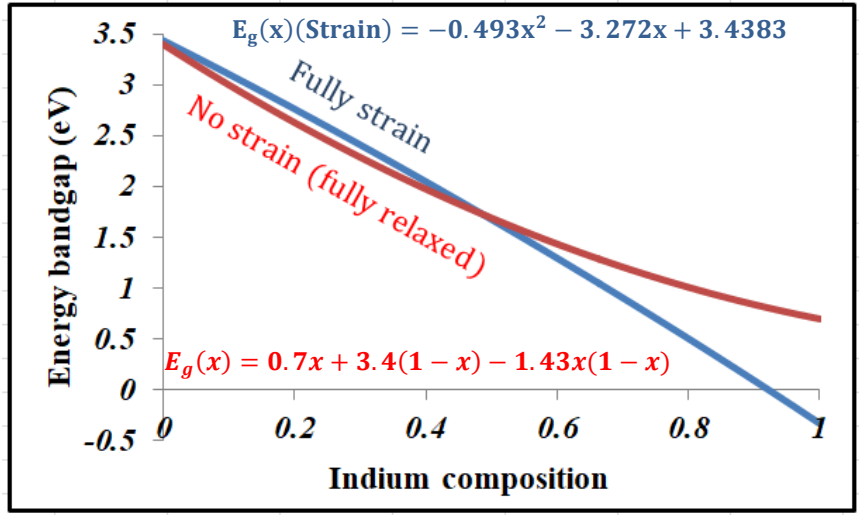

Fig 4. Energy bandgap for different compositions with (red) and without (blue) strain

By fitting a quadratic to the output of nextnano for the strained bandaps, we arrive at an analytic representation for this quantity. The equation for this fit and Vegard's law are displayed both in Fig. 4 and in Eq. 4.

$$E_g(x)(\text{Strain}) = -0.493x^2 - 3.272x + 3.4383$$
$$E_g(x)(\text{No strain}) = 0.7x + 3.4(1-x) - 1.43x(1-x) \text{ (Vegard's law)}$$
(4)

It can be seen in Fig. 4 that for strained $In_xGa_{1-x}N$ with $x > \sim 0.9$ the bandgap is negative. The concept of negative bandgap materials which are known as semi-metal materials is explained in a few reports[34], [35] and won't be discussed further here except to say that this is a potentially novel area of research for this material. Further simulations here will be restricted to $x < 0.9$.

Several theoretical models have been proposed to describe the absorption coefficient $\alpha(\lambda)$ for the InGaN ternary alloy. In Ref. [12] the following relation was proposed to parameterize the it.

$$\alpha(\lambda) = \alpha_0 \sqrt{a(x)(E - E_g) + b(x)(E - E_g)^2} \qquad (5)$$

with $E$ the photon energy, $E_g$ the bandgap, and $\alpha_0 = 10^5 \ cm^{-1}$.

The dimensionless fitting parameters, $a(x)$ and $b(x)$, are determined by fitting over the entire composition range assuming unstrained, bulk material [36].

$$a(x) = 12.87x^4 - 37.79x^3 + 40.43x^2 - 18.35x + 3.52 \tag{6}$$

$$b(x) = -2.92x^2 + 4.05x - 0.66 \tag{7}$$

Now, the energy bandgap and the absorption coefficient are available as functions of the indium composition over the entire range of the alloy. These should be combined to determine the absorption coefficient as a function of the bandgap. This was performed numerically, and the resulting parameters for the absorption coefficient as defined in Eq.5 as a function of the bandgap value is shown in Fig. 5a. In addition, the absorption coefficient vs wavelength (nm) is reported in Fig. 5b. The obtained model is valid up to 0.7 ev which is the bandgap of InN.

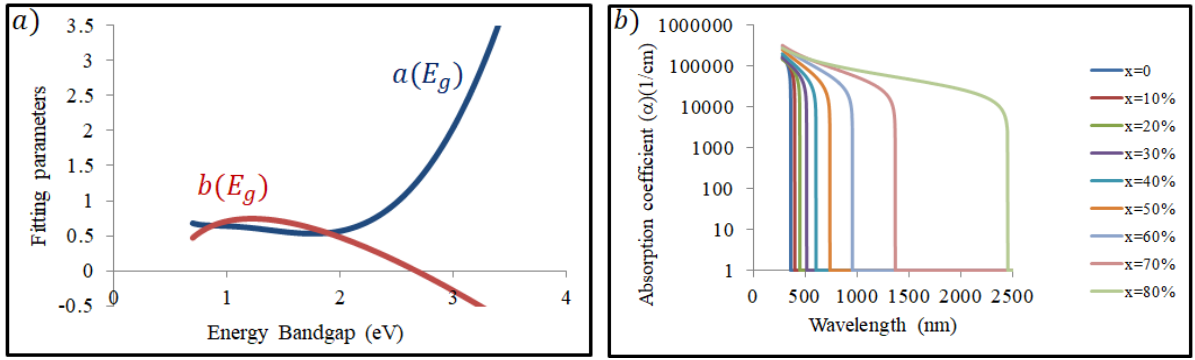

Fig 5 .a) Fitting parameters (a and b) versus energy bandgap. b) Absorption coefficient under strain for different indium alloys (0 to 80%).

Now using a polynomial fit to $a$ and $b$, we can re-write them as functions of the energy bandgap, $a(E_g)$ and $b(E_g)$.

$$a(E_g) = 0.3559 E_g^3 - 1.3645 E_g^2 + 1.5277 E_g + 0.1277 \tag{8}$$

$$b(E_g) = 0.1162 E_g^3 - 0.9968 E_g^2 + 0.19852 E_g - 0.4186 \tag{9}$$

With this model, we can predict the absorption coefficient as a function of the bandgap energy assuming either strained or unstrained material. However, because the parameterization of the absorption coefficients was derived from unstrained material with a minimum bandgap of 0.7 eV, we will limit our strained study to material with bandgap above that, i.e., from Fig. 4, $x \leq 0.70$ (strained).

Finally, the generation rates can be calculated for each layer and imported into nextnano.

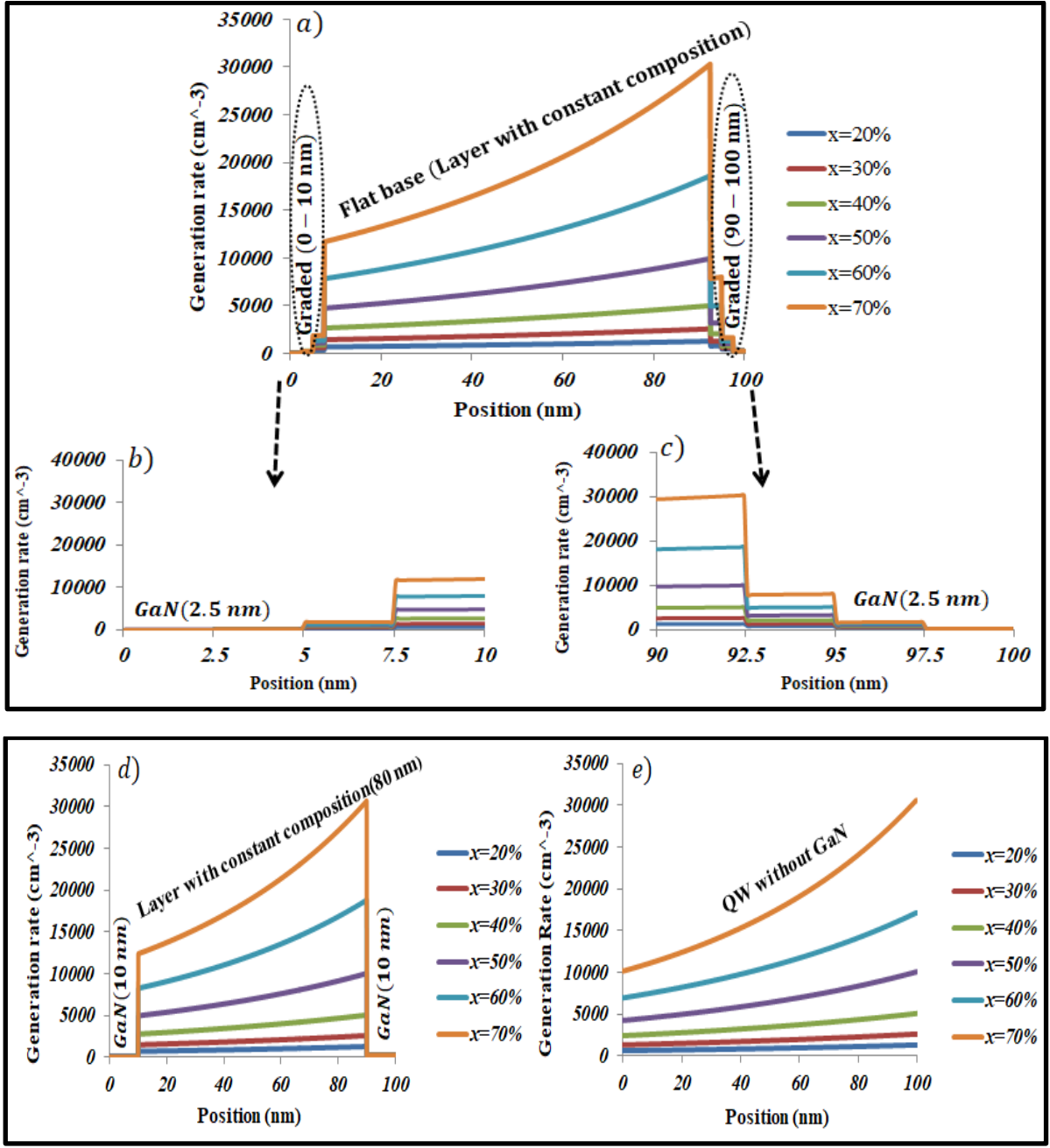

Fig6. a) Generation rate for (Structure A), b)Zoom in 0-10 nm (Structure A), c)zoom in 90-100 nm (Structure A), d) Generation rate for (Structure B), e) Generation rate for (Structure C)

The step size used to convert the graded structures to step graded was 2.5 nm which the generation rate related to that is shown in Fig6. a, b, and c. In order to understand the quality of convergence for choosing this step size, the solar efficiency and current density for two smaller step sizes (2 and 1 nm) are shown in Fig. 7. Fig. 7a compares the light $J$-$V$ plot using the three step sizes for Structure A. The open circuit voltages are virtually unchanged, however the short circuit current, Fig. 7b, and the resulting solar cell efficiency, Fig. 7c, do evolve with step size.

Generally as the step size decreases, both $J_{sc}$ and $\eta$ decrease, however they appear to be converging to values within ~1% of the reported values determined from using the 2.5 nm step size. So, we assume the 2.5 nm step size is a relatively good approximation of the values given the significant savings in computational time.

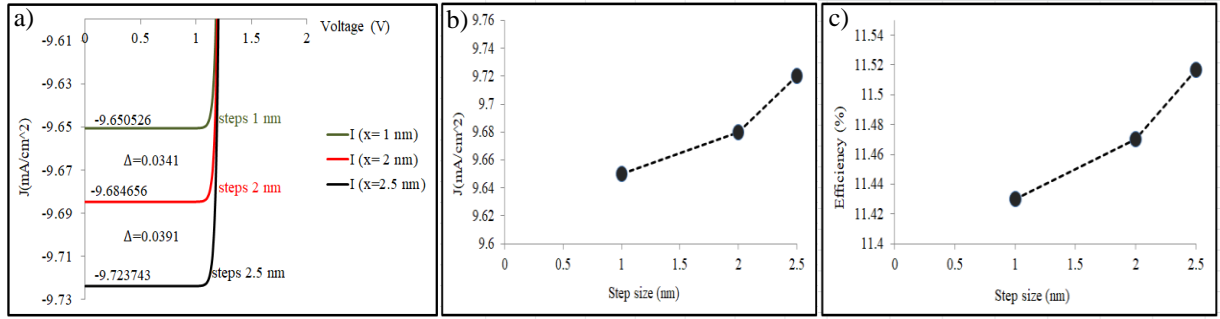

Fig.7. a) IV curve for structure A at different step sizes, b) current density at different step sizes, c) Efficiency at different step sizes

## Solar cell parameters

Now, using these generation rate profiles imported into nextnano, accurate simulations of the *J-V* characteristics can be performed under AM1.5G (100 mW/cm$^2$) standard solar illumination, with the device temperature set to 300 K. These are shown in Figs. 8a-c for all of the studied structures. From these curves the fill factors and solar cell efficiencies are determined manually.

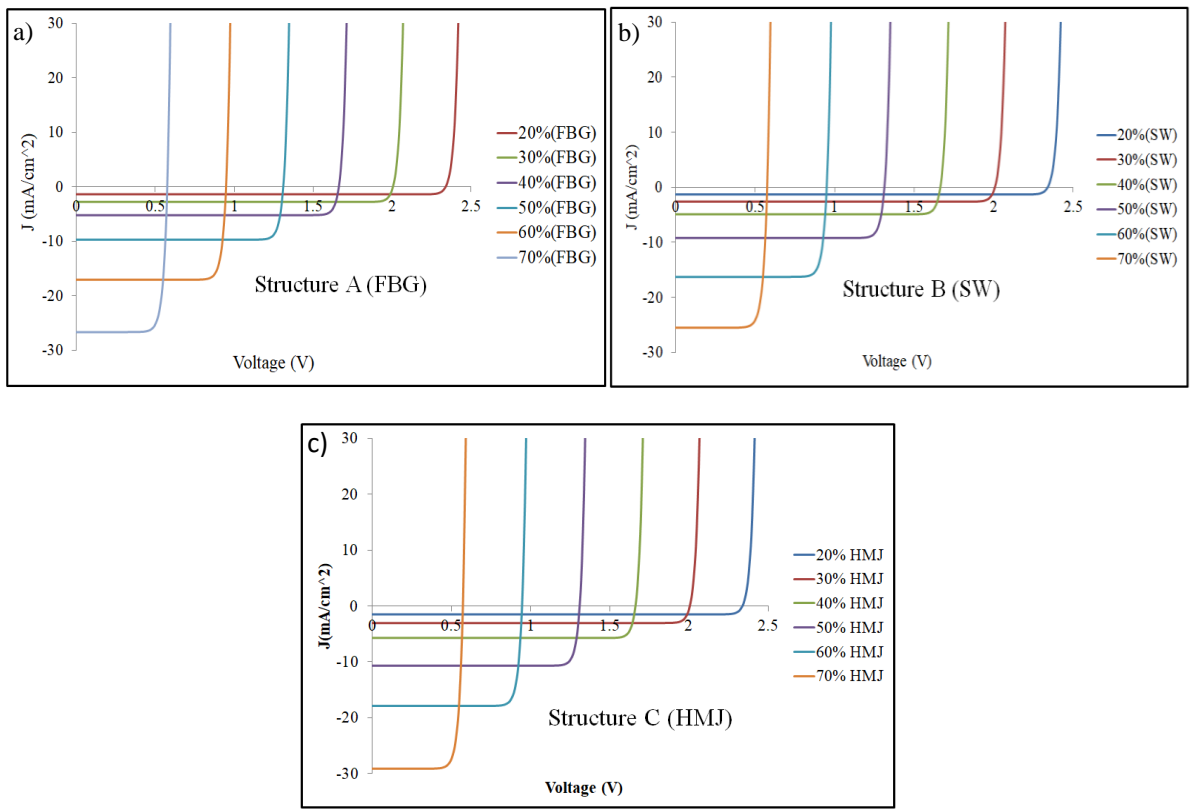

Fig.8. Illuminated IV curve for a) structure A, b) structure B, c) structure C are given.

As expected, by increasing the indium composition $V_{oc}$ generally decreases due to the overall decrease in the bandgap, as seen in Fig. 9, with the above differences between the structures. At the same time, the short circuit current, $I_{sc}$, generally increases with increasing indium composition, Fig. 10.

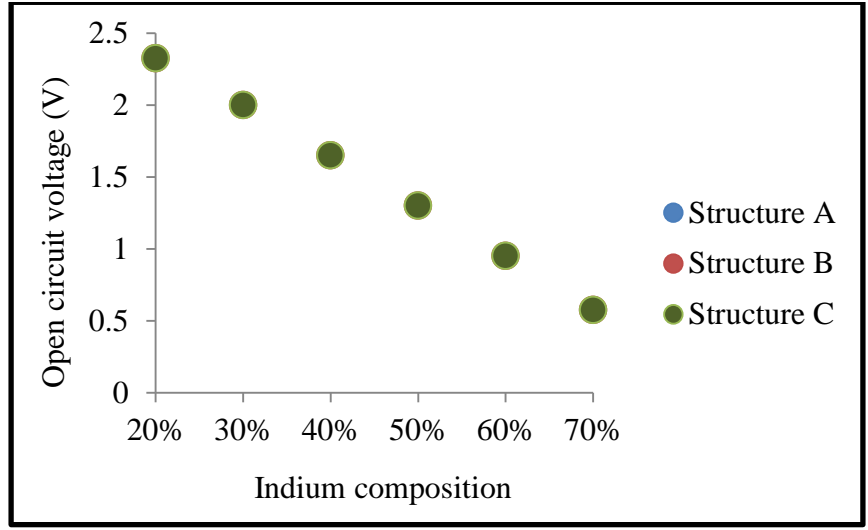

Fig 9. Open Circuit voltage for three samples

The short circuit current density, $J_0$, for the three structures is shown in Fig. 9.

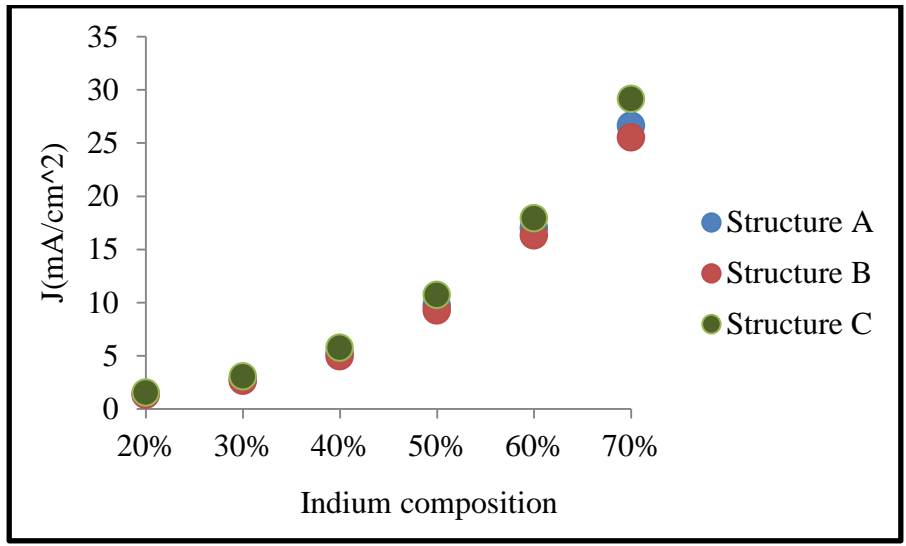

Fig.10. Short circuit current density, $J_{sc}$, for the three structures

The Fill Factor (FF) is defined as the ratio between the maximum power ($P_{max} = J_m V_m$) created by the solar cell and $V_{oc} J_{sc}$ as follows.

$$Fill\ Factor = \frac{J_m * V_m}{J_{sc} * V_{oc}} \tag{10}$$

The FF, shown in Fig.11, decreases with increasing indium composition in all structures with a maximum of ~95% at $x=20\%$, and minimum values around 80% for at $x=70\%$

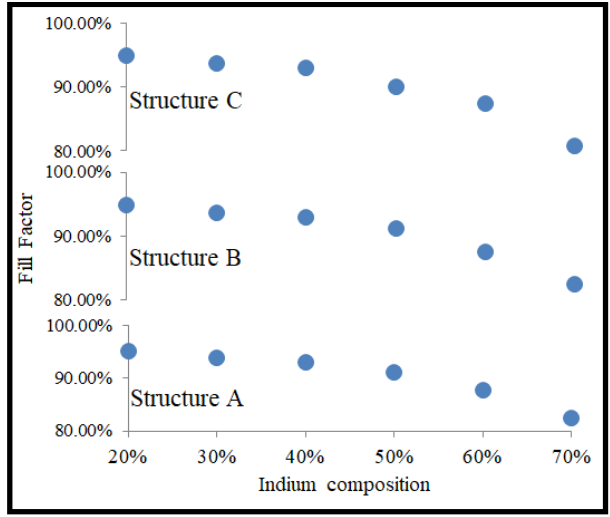

Fig.11.Fill Factor comparison for 3 samples

Finally, the solar efficiency, Fig. 12, is defined as the part of the energy in the form of sunlight that gets converted to electricity by the solar cell and is given by:

$$Efficiency(\%) = \frac{P_{max}}{P_{in}} = \frac{P_{max}}{100} \tag{11}$$

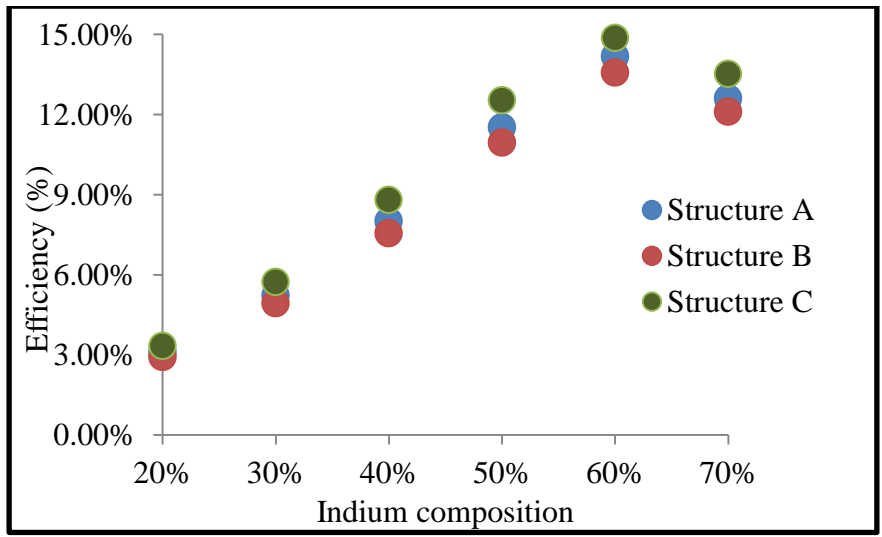

Fig.12. Solar efficiency for all samples

The solar efficiency for all structures show maximum values at 60% which is 14.18%, 13.57% and 14.86%, respectively for structures A, B, and C. This is significantly lower than some other reports due in part to the significantly thinner structure of only 100 nm [11]. In general, we have, $\eta_{Structure\ C} > \eta_{Structure\ A} > \eta_{Structure\ B}$. So, we find that structure C, a simple homojunction, provides for the most efficient solar cell. This is likely not surprising, however physically growing structure C to high quality would be challenging. However, structure A has been demonstrated to be grown at a high quality[37], while its performance is not substantially worse that structure C.

**Conclusion**

Although, at 60% indium composition the efficiency of structure C (homojunction) is larger than structure A (FBG with triangle well), there are still advantages in designing and using graded layers in the contact regions. One of the most significant advantages is the formation of the p- and n-type contact layers due to polarization doping as p-type doping in group III-V materials is extremely inefficient. Additionally, we calculated the effect of strain on the absorption coefficient and the energy band-gap, creating a Vegard's law-like relation for the bandgap of InGaN strained to a GaN substrate. This predicts a novel negative bandgap.